\documentclass[prb,superscriptaddress,aps]{revtex4}
\usepackage{graphicx}
\usepackage{dcolumn}
\usepackage{bm}
\newcommand{\be}{\begin{equation}}
\newcommand{\ee}{\end{equation}}
\newcommand{\bea}{\begin{eqnarray}}
\newcommand{\eea}{\end{eqnarray}}
\newcommand{\beb}{\begin{eqnarray*}}
\newcommand{\eeb}{\end{eqnarray*}}
\begin{document}

\title{Modified Coulomb gas construction of quantum Hall states
from non-unitary conformal field theories}

\author{M. V. Milovanovi\'{c}}
\affiliation{Institute of Physics, P.O.Box 68, 11080 Belgrade,
Serbia}

\author{Th. Jolic\oe ur}
\affiliation{Laboratoire de Physique Th\'{e}orique et Mod\`{e}les
Statistiques, Universit\`{e} Paris-Sud, 91405 Orsay, France }

\author{I. Vidanovi\'{c}}
\affiliation{Institute of Physics, P.O.Box 68, 11080 Belgrade,
Serbia}

\date{\today}

\begin{abstract}
Some fractional quantum Hall states observed in experiments
may be described by first-quantized wavefunctions with special
clustering properties like the Moore-Read Pfaffian for filling factor
$\nu =5/2$. This wavefunction has been constructed by constructing
correlation functions of a two-dimensional conformal field theory (CFT)
involving a free boson and a Majorana fermion. By considering other CFTs
many other clustered states have been proposed as candidate FQH states
under appropriate circumstances. It is believed that the underlying CFT
should be unitary if one wants to describe an \textit{incompressible}
 i.e. gapped liquid state. We show that by changing the way one derives the
wavefunction from its parent CFT it is possible to obtain an
incompressible candidate state when starting from a non-unitary
parent. The construction mimics a global change of parameters in the
phase space of the electron system. We explicit our construction in
the case of the so-called Gaffnian state (a state for filling factor
2/5) and also for the Haldane-Rezayi state (a spin-singlet state at
filling 1/2). We note that there are obstructions along this new
path in the case of the permanent spin-singlet state of Read and
Rezayi which can be characterized as a robust gapless state.

\end{abstract}
\maketitle

\section{INTRODUCTION}
It is well known that
two-dimensional electron gases in high magnetic field may form
incompressible liquids with novel properties.
This phenomenon called the fractional quantum Hall effect (FQHE)
has been studied by various theoretical methods in the past twenty years.
One successfull approach is the use of explicit trial wavefunctions (WFs)
explicitly written in the first-quantized language.
The FQHE happens when the electrons occupy the low-lying Landau levels
that are the quantized energy levels of a particle in a plane
submitted to a perpendicular magnetic field. Its appearance also requires
special commensurable ratios between the number of electrons and the number
of states in the occupied Landau level.
After the success of the Laughlin wavefunction~\cite{RBL}, the construction
of so-called ``composite fermion'' (CF) wavefunctions
 has been very successful~\cite{Jain89,Heinonen,JainBook}
in describing many of quantum Hall incompressible liquids observed to date.
However there are some states observed in experiments that do not fit
easily in this scheme, the most prominent case being a state that forms
at filling factor $\nu =5/2$. This state is beyond the reach of the previously
mentioned theories because it has an even denominator which is forbidden
in the CF constructions. There is an interesting proposal due to Moore and Read\cite{mr}
to capture the physics of this state
by introducing the notion of pairing of composite fermions.
The explicit wavefunction they propose, hereafter called the ``Pfaffian'',
has several desirable properties
like good overlap with the results of exact diagonalization of small systems
and presumably has a gap to bulk density excitation. This state has also
quasiparticle excitations with fractional charge $\pm e/4$ that have non-Abelian statistics,
an unprecedented phenomenon in physics. There are some experimental evidences~\cite{Radu08}
for these peculiar fractionally charged states. This special state
has attracted attention in the context of quantum computing~\cite{rmp}.

It is known that many of the WFs proposed in the literature can be derived
from two-dimensional massless quantum field theories possessing conformal invariance
i.e. conformal field theories (CFT). From a practical point of view
the WFs can be written as correlation functions of some operators in a given CFT.
The Laughlin wavefunction can be constructed from expectation values
of exponentials of a free massless boson. The Pfaffian of Moore and Read
can be constructed from a free massless boson and an additional massless
fermion of Majorana type (the same type of fermion that appears in the critical theory
of the classical two-dimensional Ising model).
One can also, starting from a given CFT, deduce candidate WFs that have interesting
properties inherited from its parent. For example, starting from parafermion CFT,
Read and Rezayi~\cite{Read96,Read99} have constructed WFs that have special vanishing properties~:
they are states that vanish when clusters of $k$ (bosonic) particles are at the same point.
This raises the following question~: are there \textit{a priori} restrictions on the CFTs
from which one can derive WFs ? Notably is unitarity of the CFT a necessary
condition to derive candidate WFs for incompressible states ?
non-unitary CFT appear naturally in some physical systems, for example
usual percolation has a critical point which is described by a CFT with central
charge c=-2, a non-unitary CFT. In the FQHE context, there is the so-called
Haldane-Rezayi wavefunction~\cite{hr} with filling factor $\nu =1/2$ which is derived
from a non-unitary CFT and it is known to be gapless~\cite{rg}. So it may describe
a critical point between different quantum Hall states but certainly not
a bulk incompressible FQHE state. It has been argued by Read~\cite{nick}
that, generically, non-unitary CFTs lead to compressible gapless states.
Recently many families of WFs with interesting algebraic properties
have been constructed from non-unitary CFTs~\cite{steve} so it may very well
be that they do not describe bulk gapped FQHE states.

When formulated in first-quantized language, most of the quantum Hall WFs
have expressions that do not translate easily in the Fock basis of second quantization.
In general a true FQHE WF has components on all Fock basis states allowed by symmetry.
This is the case for example of the exact eigenstates of the Coulomb problem in the LLL
as obtained by exact diagonalization. However some of the model trial WFs
have simpler expressions. It has been known for a long time~\cite{HRsqueezed} that the celebrated
Laughlin wavefunction has nonzero components only in a restricted set of the
Fock basis. Indeed one can define a partial order relation onto the Fock states
and there is a special set of occupation numbers i.e. a special basis element
that is ``greater'' than all the other terms appearing in the expansion of the Laughlin state.
This special element is called the dominant partition in the language of polynomials
of several variables (our convention in this paper) and is also called the root partition
in the literature. This means in practice that these WFs are simpler than a generic state
and also that some of their properties are encoded/can be read off the dominant partition.
So the contemplation of the dominant partition may be a tool to uncover
previously unknown relationship between quantum Hall state, to be proved by other methods
for the whole WF.

In this paper we show that the analysis of the dominant partition
suggests that some quantum Hall WFs, constructed from non-unitary
CFTs, may be ``cured'' by addition of extra quasihole-quasiparticle
excitations to produce presumably \textit{bona fide} gapped Abelian
quantum Hall states. This is in line with what we expect from a
critical theory located right at the boundary of a gapped phase~:
some perturbation/modification of it has to do with the bulk gapped
phase. Here we point out such a mechanism for two special quantum
Hall states, the Gaffnian state and the Haldane-Rezayi state. Even
this is suggestive, it remains to prove that it holds for the full
quantum Hall state. We show that a modification of the Coulomb gas
formulation precisely allows to prove that our identification holds
for the complete state i.e. not only one (important) term of the
Fock basis. This can be done by using special background charges as
introduced some time ago in the CFT literature.

In section II we show that by boundary insertions we can transform
the Gaffnian (bosonic) state at  filling factor $\nu = 2/3$ into the
Jain state of bosons at  $\nu = 2/3$  (this implies that fermionic
cases at $\nu = 2/5$  are  related in the same manner) and we
transform the Haldane-Rezayi spin-singlet state by boundary
insertions of charged excitations into a (331) multicomponent
Halperin state. The states  we obtain through these transformations
are all Abelian incompressible states. In Section III we give a
general prescription in the Coulomb gas language to show that the
correspondence we found holds not only for the dominant partition
but for the full quantum Hall WFs. We also point out at least one
case for which this scheme is probably more complex~: the permanent
state which cannot be transmuted by this mechanism into a gapped
state. In Section IV we apply the boundary insertion construction to
unitary Pfaffian state to find out whether it would transform into
an Abelian state. It is shown that the Pfaffian remains stable under
these transformations. Section V contains our conclusions. In
Appendix A, we discuss briefly the neutral excitations on top of the
Haldane-Rezayi state. In Appendix B, we explicitly derive the
Moore-Read Pfaffian WF from the Coulomb gas CFT.

\section{From gapless to gapfull states via boundary insertions}

\subsection{Quantum Hall states}
We consider WFs for electrons residing in the lowest Landau level (LLL).
In the symmetric gauge, one-body orbitals are given by~:
\begin{equation}
\label{onebody}
 \phi_m (z) =\frac{1}{\sqrt{2\pi m! 2^m}} {\mathrm e}^{-|z|^2/4},
\end{equation}
where $z=x+iy$ is the complex coordinate in the plane where electrons are confined
and the positive integer $m$ gives the angular momentum of the state~:
$L_z =m\hbar$ (we have set the magnetic length to unity).
A general N-body LLL quantum state is thus of the form~:
\begin{equation}
 \Psi(z_1,\dots , z_N)=f(z_1,\dots ,z_N) {\mathrm e}^{-\sum_i |z_i|^2/4}.
\end{equation}
In the remainder of the paper we will always omit the (universal) exponential factor.
The physics of two-dimensional electrons in the LLL is governed by the
following Hamiltonian~:
\begin{equation}
\label{FQHEH}
 {\mathcal H}=\sum_i \frac{1}{2m_b}
\left(\mathbf{p}_i+\frac{e}{c}\mathbf{A}_i\right)^2
+\sum_{i<j}\frac{e^2}{\epsilon r_{ij}},
\end{equation}
where $m_b$ is the band mass of the electron, $\epsilon$ the dielectric constant
of the host semiconductor and the distance between electrons $i$ and $j$ is
$r_{ij}=|r_i-r_j|$. In the LLL the kinetic energy is quenched and in principle one has
to diagonalize the interaction potential in Eq.(\ref{FQHEH}) in the Fock space
constructed from products of one-body states in Eq.(\ref{onebody}).
Several different schemes have been developed to understand the physics of this problem
since no general analytical solution is possible. It is feasible to diagonalize
numerically the Hamiltonian above Eq.(\ref{FQHEH}) if one considers a small number of electrons
so that the Fock space is not enormously large. This method has the advantage
of being unbiased i.e. there is no a priori hypothesis on the form the many-body states
but it is limited to a small number of electrons of the order of 12 to 15, depending
on the filling factor of the LLL one wants to study. Exact diagonalization gives
the low-lying levels as a function of the conserved quantum numbers allowed by the
geometry of the system. For example in the unbounded plane and using the symmetric gauge
for the vector potential $\mathbf{A}=\frac{1}{2}\mathbf{B}\times \mathbf{r}$
the only conserved quantity is the angular momentum along the axis perpendicular to the plane
(i.e. the $\mathbf{B}$ axis). Another successfull approach is to construct explicit
trial wavefunctions. Originally this was pioneered by R. B. Laughlin who wrote down
an explicit formula~\cite{RBL} for the wavefunction of N electrons when the filling factor
of the LLL is precisely 1/3. This Laughlin wavefunction is not an exact eigenstate
of Hamiltonian Eq.(\ref{FQHEH}), however it was shown very soon by D. Haldane
that it encompasses all the physics of the exact ground state. This demonstration
was done by comparison with data from exact diagonalization~\cite{FDM}.
This approach has been extended by Jain~\cite{Jain89,Heinonen,JainBook}
to many (if not all) fractions displaying the FQHE. The wavefunctions constructed
in this approach are known under the name of ``composite fermion'' wavefunctions.
Similarly they are not exact eigenstates of the full many-body problem
but comparison with exact diagonalization results show that they capture the essential physics.
A detailed account is given in Jain's book~\cite{JainBook}. The composite fermion
wavefunction are built from Jastrow-like correlation factors in a way that generalizes
the usual notion of Slater determinant. It is also feasible to use these wavefunctions as an
alternate basis set and to diagonalize the Hamiltonian Eq.(\ref{FQHEH}) in this basis.
This has proved useful to describe for example the fate of electrons in
quantum dots~\cite{Jeon04,Jolad09}. The same exact diagonalization techniques have been
employed also in the context of bosonic systems, motivated by the developments
of experiments on ultracold gases. This allows for example
for studies of the crystalline structures~\cite{Barberan06,Baksmaty07}
that form on small systems analogous to quantum dots in electronic systems.

Some trial wavefunctions have been also obtained by arguments based upon
conformal field theories. In this approach one construct wavefunctions by
computing expectation values of a product of operators of a definite CFT.
This is a way to reproduce the Laghlin wavefunction and it leads to many interesting proposals,
the most physcially relevant so far being the Pfaffian wavefunction of Moore and Read.
In the CFT approach, it is not known from the beginning if the wavefunction
is relevant to a given physical situation, one has to compare its predcitions
with exact diagonalization and/or experimental facts.

\subsection{Quantum Hall polynomials}
All the physics is contained in the analytic function $f$. This function
can be expanded in powers of the $z_i$ coordinates and a general term in the expansion
is characterized by the set of occupation numbers of the one-body orbitals
\{$n_m, m=0,1,2,\dots $\}. We will consider also bosonic quantum Hall states
for which one can have $n_m >1$.
If we start from a fermionic state $\Psi_F$ then antisymmetry and LLL means
that necessarily one can factor out a Jastrow-like factor~:
\begin{equation}
 \Psi_F = \prod_{i<j}(z_i - z_j) \Psi_B,
\end{equation}
with $\Psi_B$ a bosonic i.e. symmetric LLL wavefunction.
So it is enough to consider bosonic wavefunctions.
The filling factors of these two states are then related by
$1/\nu_F=1+1/\nu_B$.
A given configuration of occupation numbers $(n_0,n_1,n_2\dots )$ fully characterizes
each term of the expansion of $f$. The set of occupation numbers can be regarded
as giving a partition of N since $N=\sum_m n_m$. Alternatively
one can also specify the same configuration by giving all the $m$ values that
appear with nonzero occupation numbers $(m_1..m_1m_2..m_2\dots )$
where each $m$ is repeated $n_m$ times. This set of numbers then defines
equivalently a partition of the total angular momentum $L_z=\sum_m mn_m$.
In the physics literature it is common to specify the set of occupation numbers
while the mathematical literature~\cite{MacDoBook} on symmetric polynomials uses instead
the partitioning of $L_z$.
 A partition $\lambda$ defines also a unique symmetric monomial
$m_\lambda$ given by~:
\begin{equation}
 m_\lambda = z_1^{k_1} \dots z_N^{k_N}+{\mathrm permutations}.
\end{equation}
This can be considered as a (unnormalized) wavefunction for N bosons in the LLL
where the quantum numbers $k_i$ of occupied orbitals can associated
in a one to one correspondence to a set of occupation numbers \{$n_m$\}.
For example the monomial for N=3
$m=z_1^2 z_2 z_3 +{\mathit perm}$
is defined by the partition (0210$\dots$) since there are two bosons in the
m=1 orbital and one boson in the m=2 orbital. An arbitrary bosonic
WF in the LLL can be expanded in terms of such monomials, each of them being indexed
by a partition~:
\begin{equation}
    f =\sum_\lambda c_\lambda m_\lambda ,
\label{expans}
\end{equation}
 where $c_\lambda$ are some coefficients. For a given $f$ it may happen that not
 all partition appear in the expansion above. Indeed there is a partial ordering on partitions
called the dominance ordering~: let $\lambda$ and $\mu$ two partitions then
$\lambda \geq \mu$ if $\lambda_1+\dots+\lambda_i \geq \mu_1+\dots +\mu_i$ for all $i$.
This is only a partial order~: it may happen that the relation above does not
allow comparison of two partitions. Some of the trial wavefunctions proposed in the
FQHE literature have the property that there is a dominant partition with respect to
this special order and all partitions appearing in the expansion Eq.(\ref{expans})
are dominated by a leading one~:
\begin{equation}
 \Psi = \sum_{\mu \leq \lambda} c_\mu m_\mu .
\end{equation}
This was first noted by Haldane and Rezayi~\cite{HRsqueezed} in the
case of the Laughlin wavefunction.  This property of dominance is
also shared by many of the special orthogonal polynomials in several
variables~\cite{MacDoBook}. It was realized after the work of Feigin
et al. in Ref.(\onlinecite{Feigin}) that the Read-Rezayi (RR) trial
wavefunctions~\cite{Read96,Read99} are all particular cases of the
so-called Jack polynomials. These symmetric polynomials noted
$J^{\alpha}_{\lambda}$ are a family indexed by a partition $\lambda$
and depend upon one parameter $\alpha$. In fact we have~:
\begin{equation}
 \Psi_{RR}^{(k)}= {\mathcal S}
\prod_{i_1<j_1}(z_{i_1}-z_{j_1})^2  \dots
\prod_{i_k<j_k}(z_{i_k}-z_{j_k})^2
\propto J^{-(k+1)}_{\lambda_k}(\{z_i\}),
\end{equation}
where the first equality defines the Read-Rezayi states, one divides the particles
into $k$ packets and  ${\mathcal S}$ means symmetrization the product of partial Jastrow factors.
In the case of the RR states we have $\alpha = -(k+1)$ and
$\lambda_k =(k0k0k0\dots)$. The usual bosonic Laughlin wavefunction is the special subcase
when there is only one packet $k=1$ and the Moore-Read Pfaffian corresponds to the case $k=2$.
In general the filling factor of the order-$k$ RR state is $\nu =k/2$.
Such WFs may describe some incompressible liquids of rapidly rotating bosons
or, after due multiplication by a Jastrow factor, some elusive quantum Hall states
in the second Landau level of electrons like the $\nu=12/5$ state.

It is convenient also to study the FQHE in the spherical geometry which has no boundaries
and possesses the full rotation symmetry. In this case the LLL is finite-dimensional
since the sphere has a finite area. Basis (unnormalized) functions of the LLL can be taken as~:
$\Phi_{S}^{M}=u^{S+M}v^{S-M}, M=-S,\dots ,+S$ where
$u=\cos(\theta/2) {\textrm{e}^{i\phi/2}}$, $v=\sin(\theta/2) {\textrm{e}^{-i\phi/2}}$
and the flux through the sphere is 2S in units of the flux quantum $h/e$.
The stereographic projection from the sphere to a plane gives a one-to-one
mapping of the wavefunctions in these two geometries. When written on the sphere
quantum Hall WF have a linear relation between flux and number of particles
$2S =(1/\nu)N-\sigma$ when there is in general a nonvanishing offset $\sigma$
called the ``shift'' in the FQHE literature wrt to the defining relation of the filling factor.
On the sphere the finite number of orbitals leads to a finite set of occupation numbers
hence the dominant partition is now given unambiguously by these numbers.

Finally when considering WFs for systems with more than one component (like electrons with spin)
it is convenient to define the Halperin wavefunctions~\cite{halperin}
 with several Jastrow-like factors~:
\begin{equation}
\Psi_{mm^\prime n}=
\prod_{i,j\in A}(z_i-z_j)^m
\prod_{k,l\in B}(z_k-z_l)^{m^\prime}
\prod_{a\in A,b\in B}(z_a-z_b)^n,
\label{Halperin}
\end{equation}
where there are two components and the respective indices belong to subsets $A$ and $B$.

\subsection{The Gaffnian state}
In the fermionic Laughlin wavefunction at filling factor 1/3 any pair of particles have relative
angular momentum at least three. If we consider the projector onto relative angular momentum
one for each pair and sum these projectors then the Hamiltonian~:
\begin{equation}
 {\mathcal H_2^1} = \sum_{i<j}{\mathcal P_2^1}(ij)
\end{equation}
has a densest ground state with zero energy which is exactly the Laughlin wavefunction.
Similarly the bosonic Moore-Read Pfaffian is the densest zero-energy ground state
of the Hamiltonian defined through $P_{3}^{2}(ijk)$, excluding states where three particles
have relative momentum two. One can ask now what is the densest zero-energy state
when we consider excluding relative angular momentum three for three particles and the unique answer
is the so-called bosonic Gaffnian WF introduced originally in Ref.(\onlinecite{steve}) as a
natural generalization of the Pfaffian state. Its coordinate first-quantized expression
is~:
\begin{equation}
 \Psi_G ={\mathcal S}
\prod_{i<j\leq N/2}(z_i-z_j)^2
\prod_{N/2< k<l}(z_k-z_l)^2
\prod_{m\leq N/2 < n}(z_m-z_n)
\prod_{p\leq N/2}\frac{1}{z_p-z_{p+N/2}},
\label{lagaffe}
\end{equation}
where $\mathcal S$ stands for symmetrization.
It was recognized as the Jack polynomial $J^{-3/2}_{\lambda_G}(\{z_i\})$ with dominant partition~:
\begin{equation}
\lambda_G =(2002002\dots)
\label{gaf}
\end{equation}


While Jain wavefunctions are not in general Jack polynomials however they do
satisfy restrictive rules on the partitions that appear when expanded
in terms of monomials. Notably there is a dominant partition of Jain states
and in the case of bosons at filling $\nu = 2/3$ it is given by~:
\begin{equation}
\lambda_{2/3}=(2010110110102)
\label{jainroot}
\end{equation}
as found in References (\onlinecite{bh}) and (\onlinecite{rbh}).
By counting the number of particles and fluxes, the
relationship between the number of particles and the number of flux
quanta is the same as in the Gaffnian state~: $N_{\phi} = \frac{3}{2} N_{e} -
3$. If we introduce one extra flux quantum in the Gaffnian state without
changing the number of particles the new state may be described as an
additional zero somewhere in the configuration of the Gaffnian (i.e.
a Laughlin quasihole) \cite{bh}. A state with a more
uniform distribution of particles is obtained with a pair of
half-flux non-Abelian quasiholes, where one quasihole is put on the
North Pole and the other on the South Pole (in the sphere geometry).
This is represented by
the following partition~:
\begin{equation}
\lambda_{1qh-1qh}=(11011011011011),
\label{2qh}
\end{equation}
(compare with (\ref{gaf}) the number of flux quanta and particles).
This is the bulk configuration of Jain state (\ref{jainroot}). Due to
the same number of flux quanta and particles in (\ref{gaf}) and
(\ref{jainroot}) i.e. Gaffnian and Jain state, this suggests that the
Jain WF can be described as a Gaffnian WF with neutral
quasiparticle-quasihole excitations on the boundaries of the system. This
identification was first done in Ref.~(\onlinecite{bh}).

\subsection{The case of Haldane-Rezayi state}
The Haldane-Rezayi state was introduced in Ref.~(\onlinecite{hr}) as a
FQHE state with some kind of pairing. It is a global
spin-singlet that can be described as a collection of spin-singlet
pairs with pairing function $g(z) \sim \frac{1}{z^{2}}$ at filling
factor $\nu = \frac{1}{2}$ in the fermionic case.
\begin{equation}
 \Psi_{HR}=\sum_{\sigma\in S_{N/2}}
{\textrm{sign}}\sigma
\frac{1}{
(z_1^{\uparrow}-z^{\downarrow}_{\sigma(1)})^2
\dots
(z_{N/1}^{\uparrow}-z^{\downarrow}_{\sigma(N/2)})^2}
\prod_{i<j} (z_i-z_j)^q ,
\label{HRdef}
\end{equation}
where $q=2$.
Before realizing that Haldane-Rezayi is
a critical (gapless) state \cite{rg} there were attempts~\cite{mmnr,gn1}
 to construct the edge theory for this system on the
assumption that the HR system may represent a gapped phase even though
it is related to a non-unitary CFT. One of these attempts~\cite{gn1}
describes the edge of the HR system as the
edge of a (331) Halperin two-component state, i.e. one of the well-known
Halperin states~\cite{halperin} that are certainly gapped.

We know show that, by inspection of the dominant partitions, there is evidence
for a change of physics due to boundary insertions as we saw in the Gaffnian case.
Explicitly the dominant partition of the Haldane-Rezayi state is~:
\begin{equation}
\lambda_{HR}=(\bar{2}000\bar{2}000\bar{2}0 \cdots \bar{2}000\bar{2}),
\label{hrroot}
\end{equation}
where $\bar{2}$ means double occupancy of a \textit{single}
orbital with both spins $\downarrow\uparrow$.
On the other hand the root configuration of the (331) Halperin state is
\begin{equation}
\lambda_{(331)}=(XX00XX00XX \cdots XX00XX)
\label{331root}
\end{equation}
where $XX$ stands for $\uparrow \downarrow + \downarrow \uparrow$, i.e.
 a symmetric superposition of the neighboring opposite spins~\cite{sky}.

It is important to note, by examining (\ref{hrroot}) and
(\ref{331root}), that for the same number of electrons there is one
more orbital in the (331) case as can be expected by comparing flux
and particle number relations in the Haldane-Rezayi case: $N_{\phi}
= 2 N_{e} - 4$, and in the $(331)$ case: $N_{\phi} = 2 N_{e} - 3$.
The extra flux quantum can be introduced in the Haldane-Rezayi
state as an Abelian Laughlin quasihole and therefore as an extra
zero~\cite{bh} in (\ref{hrroot}) or as two non-Abelian
quasiholes~\cite{comment} symmetrically at the boundaries of the system as in
(\ref{331root}) . Therefore this suggests again that the (331) Halperin state
can be derived by
insertions of global non-Abelian excitations in a ``parent'' Haldane-Rezayi
state. The case of neutral excitations is similar and is discussed in Appendix A.

\subsection{Discussion}
In this section we have shown that the special relationship via boundary
excitations between Gaffnian and other $W_{k}(k + 1, k + r)$
generalizations~\cite{bgs} at $\nu = \frac{k}{r}$ and Jain states
of bosons at  $\nu = \frac{k}{r}$ as demonstrated in Ref.(\onlinecite{rbh}) is
not unique but extends to other non-Abelian gapless states like HR
i.e. those states connected to non-unitary CFTs.
These states are presumably at a phase boundary
to a gapped FQHE state. Tweaking of interactions or imposing global change
like with a magnetic field parallel to the 2D electron gas plane may
lead the system from the critical point described by the non-Abelian
gapless state into a stable Abelian gapped state and phase.
In the next section we use the Coulomb gas formulation to extend our argument
beyond merely the dominant partition of the monomial expansion to the full WF.

\section{Boundary insertions in the language of Coulomb gas correlators}

\subsection{CFT formalism and FQH states}
A bulk quantum Hall fluid is an incompressible liquid which
is spatially featureless. When sitting on a sphere it will spread out
to form a uniform film that is invariant by the rotation group acting upon the sphere.
The corresponding quantum state should thus be annihilated by all
the generators of the rotation group~:
\begin{equation}
 L^{+} \Psi =  L^{-} \Psi = L_{z} \Psi = 0..
\end{equation}
The spherical geometry is of course a purely theoretical construct. We can
translate these conditions on the realistic planar geometry by using
the stereographic projection. The rotation operators are then differential operators
acting upon the particle coordinates~:
\begin{equation}
L^{+} = E_{0},
\quad L^{-} = N_{\phi}\sum_{i = 1}^{N} z_{i}  - E_{2},
\quad L_{z}
= \frac{1}{2} N N_{\phi} - E_1,
\quad  {\textrm{where}}\quad
E_{n} =  \sum_{i = 1}^{N} z_{i}^{n} \frac{\partial}{\partial
z_{i}}
\end{equation}
If we suppose that the WF is given by a correlation function
of some operators of a quantum field theory
then we have the following conditions~:
\begin{eqnarray}
&&\sum_{i = 1}^{N} \partial_{i} \langle 0|\phi_{1}(z_{1}) \cdots
\phi_{N}(z_{N})|0\rangle = 0, \nonumber \\
&&\sum_{i = 1}^{N} (z_{i} \partial_{i} + h_{i})\langle
0|\phi_{1}(z_{1}) \cdots \phi_{N}(z_{N})|0\rangle = 0, \nonumber \\
{\rm and} \nonumber \\
 &&\sum_{i = 1}^{N} (z_{i}^{2} \partial_{i} + 2 z_{i}
h_{i})\langle 0|\phi_{1}(z_{1}) \cdots \phi_{N}(z_{N})|0\rangle = 0.
\end{eqnarray}
These are the conditions for invariance under the global conformal group in two dimensions.
It is thus clear that any CFT which is by definition invariant under
the larger local conformal symmetry group will satisfy these conditions.
In a given CFT the fields $\phi_{i}$ are the (quasi)primary fields and $h_{i}$ are the
corresponding conformal weights.
Some quantum Hall WFs can be derived fromn correlators of operators of two-dimensional
massless quantum field theories, the example of the Moore-Read Pfaffian is given in Appendix B.

\subsection{The ``Gaffnian'' state}
The Gaffnian WF is built form the minimal\cite{steve,eddy} model ${\cal M}_{2}(3,5)$
The central charge is this non-unitary CFT is~:
\begin{equation}
c = \frac{r (k - 1)}{k + r} (1 - k(r - 2)) = - \frac{3}{5}.
\end{equation}
One way to construct this CFT and its correlators  is to start from a free
boson theory and introduce a background charge~\cite{gin,gn2,df,zub,car}
by adding an extra term to the energy-momentum tensor~:
\begin{equation}
T(z) = - \frac{1}{2} :\partial x(z) \partial x(z): + i \sqrt{2}
\alpha_{0} \partial^{2} x(z),
\end{equation}
where the free boson is field $x(z)$. This additional contribution
leads to a central charge~:
\begin{equation}
c = 1 - 24 \alpha_{0}^{2}.
\end{equation}
One should then think~\cite{gin} of the background
charge $ - 2 \alpha_{0}$ as being ``at infinity''.
In the FQHE formulated on the sphere this means simply that
the charge is located at the pole of the sphere which is sent to infinity
by stereographic projection.
The only non-vanishing
correlators in the case of 2-point function are~:
\begin{equation}
\langle V_{\beta}(z) V_{2 \alpha_{0} -\beta}(w)\rangle = \frac{1}{(z
- w)^{2 \beta (\beta - 2 \alpha_{0})}},
\end{equation}
where the vertex operators are given by~:
\begin{equation}
V_{\beta}(z) = :\exp(i \sqrt{2} \beta x(z)):.
\label{vertexo}
\end{equation}
These two operators $V_{\beta}$ and $V_{2 \alpha_{0} -\beta}$
are adjoint to each other and their conformal weight is
$h = \beta (\beta - 2 \alpha_{0})$.
In our case we want $ 1 - 24 \alpha_{0}^{2} = - \frac{3}{5}$ so that
$\alpha_{0} = \frac{1}{\sqrt{15}}$.

We know that
the non-Abelian quasihole derived from the Gaffnian state is described by a
product of a field $\sigma$ of the minimal model ${\cal M}_2 (3,5)$
(the neutral part) and a bosonic
vertex operator (the charge part).
The field
$\sigma$ has a conformal weight equal to $h_{\sigma} = -
\frac{1}{20}$. The corresponding values of $\beta$'s in the bosonic
representation are thus~:
\begin{equation}
\beta (\beta - 2 \alpha_{0}) = - \frac{1}{20} \rightarrow
\end{equation}
\begin{equation}
\beta_{1,2} = \alpha_{0} \pm \sqrt{\alpha_{0}^{2} - \frac{1}{20}} =
\frac{1}{\sqrt{15}} (1 \pm \frac{1}{2}).
\end{equation}
The vertex operator ``at infinity'' is~:
\begin{equation}
V_{\beta_{0} = - 2 \alpha_{0}} = :\exp(-i\; 2 \;\sqrt{2}
\;\frac{1}{\sqrt{15}}\; x(z = \infty)):.
\end{equation}
It appears as an additional insertion in correlation functions~:
\begin{equation}
\langle V_{\beta_{0}}(z = \infty)\;\;\; \cdots \;\;\;\rangle.
\end{equation}
We can recover the ordinary bosonic theory if we insert vertex
operators with $\beta = - \frac{\beta_{0}}{2}=\alpha_0$ at  two ends
- $z = \infty$ and $ z = 0$ i.e. the two poles of the sphere in the
following way~:
\begin{equation}
\langle V_{\frac{\beta_{0}}{2}} (z = \infty)\;\; \cdots\;\;
V_{-\frac{\beta_{0}}{2}} (z = 0)\rangle,
\end{equation}
or~:
\begin{equation}
\langle V_{\beta_{0}} (z = \infty) \sigma \sigma^{\dagger}(z =
\infty)\;\; \cdots \;\;\sigma \sigma^{\dagger}(z = 0)
\rangle,\label{insertions}
\end{equation}
where we have defined~:
\begin{equation}
\sigma = :\exp( i \sqrt{2}\; \frac{3}{2} \;\frac{1}{\sqrt{15}} \;
x(z)):
\quad
{\textrm{and}}
\quad
\sigma^{\dagger} = :\exp(- i \sqrt{2}\; \frac{1}{2}\;
\frac{1}{\sqrt{15}} \; x(z)):.
\end{equation}
These two operators are related to the non-Abelian quasiparticle~\cite{footnote}.
In Eq.(\ref{insertions}) we introduce a quasihole excitation $\sigma$ through the vertex operator
Eq.(\ref{vertexo}) with exponent $\beta_{1} = \frac{1}{\sqrt{15}} ( 1 + \frac{1}{2})$.
The second vertex operator that we use for the quasihole has exponent
$\beta_{2} = \frac{1}{\sqrt{15}} ( 1 - \frac{1}{2} ) > 0$
and the same conformal dimension.
We construct the quasiparticle excitation $\sigma^{\dagger}$ through
vertex operator with exponent  $- \beta_{2}$.

The most important implication of the boundary insertions in
CFT correlators is that by additional neutralizing background charges
we recover a standard bosonic
description without background charges usually associated with Abelian FQHE states.
Indeed we have~:
\begin{equation}
\langle \exp(i \sqrt{2} \beta \; x(z)) \exp(- i  \sqrt{2} \beta \;
x(w)) \rangle_{\textrm{with neutralizing insertions}} \sim \frac{1}{(z - w)^{2
\beta^{2}}},
\end{equation}
as in the usual Coulomb gas formulation.
In the Gaffnian case, though we can not reproduce the full wavefunction
of the Jain state, we note that the neutral part of the state we obtain
can be considered as a spin-singlet state of ``spinons'' i.e. excitations~\cite{bs} created
by vertex operators with $\beta = \pm \frac{1}{2}$ .
Thus the usual correlator of the neutral Coulomb gas can reproduce a Halperin
(221) state of bosons that is closely related to the Jain state at $\nu = \frac{2}{3}$
(they share the same low-energy description~\cite{nrah}).
We find that the dominant partition of this (221) state is
\begin{equation}
\lambda_{(221)}=(XX0XX0XX0XX0XX)
\end{equation}
in the notation of section II, to be compared with the
bulk pattern of Jain state in Eq.(\ref{jainroot}).

Finally we mention that
this construction with background charges can be
generalized to other $\nu = \frac{k}{r}$ cases deduced from CFTs
associated with $W_{k}(r + 1, r + k)$ algebras using their
multicomponent Coulomb gas representations.

\subsection{Haldane-Rezayi state}

In the case of Haldane-Rezayi state~\cite{hr}
the CFT has central charge $c = -2$, it is a non-unitary ``scalar fermion"
theory~\cite{mmnr}. We now use the Coulomb gas mapping
established for this non-unitary ghost system in Ref.~\onlinecite{frie}.
For the background charge $q= - 2 \alpha_{0}$ we should have~:
\begin{equation}
1 - 24 \alpha_{0}^{2} = - 2,
\end{equation}
hence we have~:
\begin{equation}
\alpha_{0} = \frac{1}{\sqrt{8}} = \frac{1}{2 \sqrt{2}}
\end{equation}
The $\sigma$ - field needed for the neutral part description of the
non-Abelian quasihole, has conformal weight~:
\begin{equation}
h_{\sigma} = - \frac{1}{8},
\end{equation}
Therefore we have~:
\begin{equation}
\beta_{1,2} = \alpha_{0} \pm \sqrt{\alpha_{0}^{2} - \frac{1}{8}} =
\alpha_{0} = \frac{1}{2 \sqrt{2}}.
\end{equation}
The background charge is given by the insertion of the
following vertex operator ''at infinity``~:
\begin{equation}
V_{\beta_{0} = - 2 \alpha_{0}}(z) = \exp(- i \; x(z)).
\end{equation}

We implement the $\sigma$-field as~:
\begin{equation}
\sigma = \exp( i \sqrt{2} \; \frac{1}{2 \sqrt{2}}\; x(z)) = \exp(i
\frac{1}{2}\; x(z)).
\end{equation}
Therefore to recover a usual bosonic theory we can insert $\sigma$ operators
at two ends, $z = \infty$ and $z = 0$, in the following manner~:
\begin{equation}
\langle V_{\beta_{0}}(z = \infty) \sigma(\infty) \;\; \cdots \;\;
 \sigma(0) \rangle. \label{ansatz}
 \end{equation}

This parallels the boundary insertion relationship we found in the
previous section that led to the (221) state that is naturally described in
the Coulomb gas formalism. If we use the neutral fermion field
instead of the $\sigma$-field operator
we have
$\beta_{1,2} = \alpha_{0} \pm \sqrt{\alpha_{0}^{2} + 1} = \frac{1}{\sqrt{8}} (1 \pm 3)$
and again by ``trivial insertions" of a single field on both ends ( i.e.
trivial because $\beta_{1} + \beta_{2} = 2 \alpha_{0}$ is always satisfied)
we obtain again an insertion ansatz in the CFT formalism that leads to
an Abelian state described by ordinary Coulomb gas formalism.
This state should be closely related to the hierarchy/Jain's spin-singlet state at
filling factor 1/2
although we have not yet been able to find the precise relationship.

Related to this is a comment we want to make that according to (a)
what we found about the root configuration of HR state i.e. how
complex its definition is, and (b) that the neutral part of the HR
state can be decomposed into a product of Cauchy determinant and
permanent, the CFT associated with the HR state may be more general
than a single ``scalar fermion" theory. This would imply more than
one Coulomb gas necessary to describe the  neutral sector of the
state and its excitations, which is quite expected given that the
Coulomb gas description of the neutral part of the (boundary
insertions related) hierarchy and Jain spin-singlet state at
$\frac{1}{2}$ requires two Coulomb gases. (The $K$ matrix of these
states is a $3 \times 3$ matrix \cite{mmnrss}.) Nevertheless a
single ``scalar fermion" theory is, as we already seen, able to
capture the basic mechanism of neutralization that is at work in
this case.

\subsection{The permanent state}
The physics of the so-called permanent state was first described in
Ref.~(\onlinecite{Read96}). This spin-singlet state is defined in the case of
electrons at filling factor one. The state contains one power of the
Laughlin-Jastrow factor (which is the Vandermonde determinant)
and has also a BCS-like pairing part
with a pairing
function is $ \frac{1}{z}$. It can be written as ~:
\begin{equation}
 \Psi_{per}=\sum_{\sigma\in S_{N/2}}
\frac{1}{
(z_1^{\uparrow}-z^{\downarrow}_{\sigma(1)})
\dots
(z_{N/1}^{\uparrow}-z^{\downarrow}_{\sigma(N/2)})}
\prod_{i<j} (z_i-z_j)^q ,
\label{perm}
\end{equation}
where $q=1$
This is the densest zero-energy state of the projector that penalizes
the closest possible approach of three spin-1/2 particles for total spin 1/2.
We find by direct expansion of Eq.(\ref{perm}) and examination of the terms that the dominant partition
 of the permanent state is~:
\begin{equation}
\lambda_{per}=(\bar{2}0\bar{2}0\bar{2}0\bar{2}0\bar{2}\dots)
\label{permroot}
\end{equation}
in the notation of section II.
The CFT that corresponds to this permanent state is the so-called $
\beta, \gamma $ (non-unitary) commuting spinor ghost system.
It is explained in Ref.~(\onlinecite{frie}) that the
ghost system allows a representation by \textit{two} Coulomb gases. Only
one of them needs a background charge and represents a pair of
``scalar fermions" as in the CFT formalism for the HR state.
The boundary condition changing field $\sigma$ (or the spin-field) can
be represented by a vertex operator of a Bose field that does not
need a background charge. Therefore the insertions of this
field $\sigma$ at the ends of a general correlator  do \textit{not} lead
to a complete neutralization of the background. Thus, since the $\sigma$ field
in the case of permanent CFT generates a
non-Abelian excitation, its insertions on the boundaries of the permanent
system cannot lead to an Abelian gapped state contrary of the HR state.
Indeed this can be guessed already from the partition analysis~:
the insertions will transform (\ref{permroot}) into a dominant partition
 of the Halperin (111) state,
\begin{equation}
(1111111111)
\end{equation}
i.e. a dominant partition of a state that is known~\cite{WenZee} to
be gapless.


\section{Boundary insertions and the unitary Pfaffian case}

\subsection{Introduction}

The examples we have given for quantum Hall states connected to
non-unitary theories are known in the literature as critical states
- see Ref.\onlinecite{steve} for the case of Gaffnian and Ref.
\onlinecite{rg} for the Haldane-Rezayi case. They are recognized to
be at the phase boundary to the Abelian states that we described
here via boundary insertions.  Therefore our construction has the
following physical interpretation - it tells whether and in what
manner a quantum Hall state connected to a non-unitary CFT and
therefore gapless can be transmuted, via some global change of
parameters described by boundary insertions, into unitary theory
with Abelian braiding properties of excitations. Then the natural
question to ask is what happens if we apply boundary insertions to
unitary non-Abelian states; whether they will be transmuted, if the
neutralization of the background charges is complete while using CFT
constructions, into Abelian unitary states. If they are ``immune"
that would give an insight into a stability of a particular state
and a stability of its non-Abelian property. In the following we
will discuss the effect of the boundary insertions on the Moore-Read
Pfaffian state.

\subsection{The case of Pfaffian}

The Pfaffian can be built from the ${\cal M}_2 (3,4)$ minimal model
or Ising CFT. The central charge is $c = 1/2$, and this is the
simplest unitary theory which as a minimal model can be represented
in the Coulomb gas formalism. Then it is not hard to repeat the
algebra as in the Gaffnian case in Section III B to find $\alpha_{0}
= \frac{1}{2\sqrt{12}}$ and corresponding $\beta$'s for the
non-Abelian quasihole field $\sigma$ are $\beta_{1} = 3 \alpha_{0}$
and $\beta_{2} = - \alpha_{0}$. Therefore in this case it is
impossible to introduce a quasiparticle vertex operator reversing
the sign of $\beta_{1}$ or $\beta_{2}$ and achieve the
neutralization of the background charge by two
quasiparticle-quasihole pairs like in the Gaffnian case.

Therefore we established that the Pfaffian state is stable wrt
global insertions of quasiparticle-quasihole pairs; insertions will
not lead to an Abelian state. Nevertheless we should also examine ``
trivial insertions" (see below Eq.(\ref{ansatz})) i.e. those that
are made by placing a single field on both ends of the system. By
doing this we may be still just in an excited sector of the
non-Abelian theory but as we saw in the Haldane-Rezayi case we may
as well enter or make a space for an Abelian theory ((331) in the
Haldane-Rezayi case). The CFT construction can not give us an answer
for that and we have to resort to examining root configuration that
correspond to this kind of trivial insertions, to see if the outcome
may be an Abelian state. The basic root configuration of the
fermionic Pfaffian at $\nu = 1/2$ is
\begin{equation}
\lambda_{pf}=(11001100110011). \label{pffroot}
\end{equation}
Insertions of a neutral fermion on both ends would lead to the
following $L = 0$ state:
\begin{equation}
\lambda_{nf}=(10110011001101) \label{nffroot}.
\end{equation}
The bulk configuration did not change and we do not have a reason to
believe that this structure can be connected to an Abelian state.
Even if we start with two-component picture of the structure that is
ensuing after the neutralization (we end up with two Coulomb gases -
compare the discussion in III B and the relationship between
Gaffnian, (221), and Jain's atate) this will not take us out of
Pfaffian. Namely the root in Eq.(\ref{nffroot}) can be related to
the root configuration of two component (331) state but its
(anti)symmetrization leads back to Pfaffian.

Next we can consider putting non-Abelian quasiholes on two ends of
the system. The corresponding root configuration in the fermionic
Pfaffian case is
\begin{equation}
\lambda_{qh}=(101010101010101). \label{pfqhroot}
\end{equation}
If we again invoke the two-component interpretation that the CFT
allows, the bulk configuration of the root in Eq.(\ref{pfqhroot})
can be related to the bulk configuration of the root of the Jain
state, $\chi_{1,1} \chi_{2} \chi_{1}$ in the usual Jain notation, at
$\nu = 1/2$ as described by Eqs. (\ref{jrootup}) and
(\ref{jrootdown}). The state  can be rewritten as
\begin{equation}
\frac{\chi_{1,1} \chi_{1} \chi_{2} \chi_{1}}{\chi_{1}}.
\end{equation}
$\chi_{1,1} \chi_{1}$ is nothing but a (221) state which under
appropriate inclusion of derivatives and a symmetrization procedure
can be transformed into the Jain bosonic state, $\chi_{2} \chi_{1}$.
Therefore this case, with non-Abelian quasihole insertions, is
non-trivial in the sense that it might lead to a non-Abelian
composite Jain state \cite{nonAjain}:
\begin{equation}
\frac{ (\chi_{2} \chi_{1})^2}{\chi_{1}},
\end{equation}
but again non-Abelian which shows how the Pfaffian physics at $\nu
=1/2$ is immune to abelianization but can be transmuted by changing
parameters of the system into another non-Abelian state.

It is interesting to note that the (bosonic) Read-Rezayi states at
$\nu = \frac{k}{r}\; ; \; r = 2, k = 3,4$ allow the abelianization
by quasiparticle-quasihole pairs as we described in the Gaffnian
case (Section III B). This is not surprising given that these
constructions can be considered at the same time with some hierarchy
(Abelian) constructions as viable candidates for corresponding
filling factors.


\section{Conclusion}

We have shown how to construct an Abelian gapped FQHE state starting
from a FQHE state deriving from a non-unitary CFT. This construction
is done in the Coulomb gas language by the introduction of some
background charges. Since we expect that states constructed from a
non-unitary CFT are gapless it means that we have a way to construct
a gapped Abelian state whose boundary in some parameter space
presumably contains the gapless state. It is interesting to note
that the Abelian/non-Abelian character of the states is not
preserved~: in the two examples discussed in this paper the gapped
state is Abelian while it is the critical theory which is
non-Abelian. Of course the non-Abelian character of a gapless theory
is a bit formal since it is not possible to perform an adiabatic
exchange of excitations to obtain their braiding properties~: there
is no adiabatic limit since there is no gap to protect the
excitations.

While we have treated in some detail the case of the Gaffnian and the HR states
the Coulomb gas construction shows that it is more general.
However it cannot be completely general. Indeed we have an example, the permanent
state, for which this construction  is impossible with non-Abelian quasihole insertions.
It would be interesting to have a clearer understanding of this special case.

Finally we applied the boundary insertion ansatz to the unitary
Pfaffian case. The Pfaffian character and non-Abelian behavior
remain preserved under boundary insertions pointing out to the
stability of this state in the context of the FQHE of polarized
electrons at $\nu = 1/2$.

\section{acknowledgments}
 We acknowledge ``Pavle Savi\'c" grant and Partnership Hubert
Curien. M.~V.~M. and I.~V. were supported by Grant No. 141035 of the
Serbian Ministry of Science. Part of this work has been done at
KITP Santa Barbara during the workshop ``Low dimensional electron
systems".

%
%

\appendix
\section{}

We now ask whether boundary insertions can be done
in the HR state while keeping flux and particle
relation fixed i.e. in a neutral way, to transform the HR state
into a gapped state. The basic neutral excitations of the HR system
are neutral fermions and they carry only a spin
degree of freedom. After an inspection of which partitions
with boundary insertionsare still uniform (L = 0) states we
conclude that the following dominant partition~:
\begin{equation}
\lambda_{N}=(\bar{2}00\uparrow 0 \downarrow 0\uparrow 0 \downarrow 0\uparrow 0 \downarrow 00\bar{2}),
\label{nfhr1}
\end{equation}
together with the configuration that we find by exchanging ups and downs:
\begin{equation}
\lambda_{N^\prime}=(\bar{2}00\downarrow 0 \uparrow 0\downarrow 0 \uparrow 0\downarrow 0 \uparrow 00\bar{2}),
\label{nfhr2}
\end{equation}
describes the neutral fermion insertions~\cite{nregnault}.

On the other hand from (a) the study of paired fermion states~\cite{rg} and (b) the work
on spin-singlet hierarchy~\cite{mmnrss} and possible spin-singlet candidates~\cite{jain} at fillings
$\frac{1}{q}$, $q$ even,  we know that there is an
Abelian incompressible phase closely connected with the HR state.
In the hierarchy picture this is a spin-singlet state that can be
constructed by condensing spinless quasielectrons on the top
of the Halperin (332) state at $\nu = \frac{2}{5}$. We will use the expression of
the state in the Jain picture~\cite{jain}~:
\begin{equation}
\Psi = \chi_{1,1} \chi_{2} \chi_{1},
\end{equation}
where we used the usual Jain notation for $\chi_{1}$, the Jastrow-Laughlin
factor for the filled LLL  (Vandermonde determinant), $\chi_{2}$ as the wave function for
two filled LLs of all particles and $\chi_{1,1}$ as the filled LLL of both spins i.e. (110) state in the
Halperin notation. The spinless part of the wave function $(\chi_{2} \chi_{1})$ is
 the Jain state at $\nu = \frac{2}{3}$ for bosons which the dominant partition is in Eq.(\ref{jainroot}) i.e.
\begin{equation}
\lambda_{2/3}=(2010110110102).
\label{jain2}
\end{equation}
Inserting the fluxes that carry spin by $\chi_{1,1}$, after a little inspection we find~:
\begin{equation}
\lambda_{1/2}=(\bar{2}00\uparrow 0 \downarrow 0\uparrow 0 \downarrow
0\uparrow 0 \downarrow 00\bar{2})\label{jrootup}
\end{equation}
and
\begin{equation}
\lambda_{1/2}^\prime=(\bar{2}00\downarrow 0 \uparrow 0\downarrow 0
\uparrow 0\downarrow 0 \uparrow 00\bar{2})\label{jrootdown}
\end{equation}
as the basic configurations that describe the Jain state at $\frac{1}{2}$.
By comparing what we
found out about neutral fermion constructions in the HR state (Eqs. (\ref{nfhr1})
and (\ref{nfhr2})) we conclude that this
Jain state at $\frac{1}{2}$ can be realized by implementing boundary insertions of neutral fermions
in the HR state, at least when considering dominant partitions.

%
%

\section{}
The Moore-Read state is given by~:
\begin{equation}
\Psi_{MR}=\prod_{i < j} (z_{i} - z_{j})^{m} Pf(\frac{1}{z_{i} - z_{j}}),
\end{equation}
where~:
\begin{equation}
 Pf(\frac{1}{(z_{i} - z_{j})}) = \sum_{\sigma \in S_{N}} sgn\;
 \sigma \;
 \frac{1}{(z_{\sigma(1)} - z_{\sigma(2)})} \cdots \frac{1}{(z_{\sigma(N - 1)} -
 z_{\sigma(N)})},
\end{equation}
and we have a pairing part (Pfaffian) or neutral part that
corresponds to a correlator of $N$ Majorana fermion fields. The
Laughlin part or charge part is a correlator of special bosonic
vertex operators with a background charge~\cite{mr}. Explicitly for
the Pfaffian part~:
\begin{equation}
\Psi_{Pf} = Pf(\frac{1}{(z_{i} - z_{j})}),
\end{equation}
we have $h_{i} = h = \frac{1}{2}, \forall i$ that is (in the
previous notation) $(E_{2} + Z) \Psi_{Pf} = 0$ and $(E_{1} + N
\frac{1}{2}) \Psi_{Pf} = 0$ so that $M = - N \frac{1}{2}$ i.e.
$N_{\phi} = - 1$, and for the Laughlin part~:
\begin{equation}
\Psi_{L}^{m} = \prod_{i <j} (z_{i} - z_{j})^{m}
\end{equation}
the correlator is given by~:
\begin{equation}
\langle \exp(- i \; N \sqrt{m}\; \Phi(\infty)) \exp( i \; \sqrt{m}\;
\Phi(z_{1})) \cdots \exp( i \; \sqrt{m}\; \Phi(z_{N})) \rangle,
\end{equation}
for a boson field $\Phi$ and with the background charge at $ z =
\infty$, which (as we will explain more later) shifts the value of
the conformal weight of $\exp( i \; \sqrt{m}\; \Phi(z))$ from
$\frac{m}{2}$ to $ \frac{m}{2} - \frac{m}{2} N$ so that: $(E_{2} + Z
(m - m N)) \Psi_{L}^{m} = 0$ and $(E_{1} + \frac{N}{2} (m - m N))
\Psi_{L}^{m} = 0$ i.e. $M = \frac{m N (N - 1)}{2}$ and $N_{\phi} = m
(N - 1)$. Together, $\Psi_{Pf}$ and $\Psi_{L}^{m}$ lead to the Moore-Read
WF with $N_{\phi} = m (N - 1) - 1$ as expected.




\begin{thebibliography}{99}

\bibitem{RBL}
R. B. Laughlin,
Phys. Rev. Lett. \textbf{50}, 1395 (1983).

\bibitem{Jain89}
J. K. Jain,
Phys. Rev. Lett. \textbf{63}, 199 (1989).

\bibitem{Heinonen}
\textit{Composite Fermions : A Unified View of the Quantum Hall Regime},
edited by O. Heinonen (World Scientific, Singapore, 1998).

\bibitem{JainBook}
\textit{Composite Fermions},
J. K. Jain (Cambridge University Press, Cambridge, 2007).

\bibitem{mr}
G. Moore and N. Read,
Nucl. Phys. B {\bf 360}, 362 (1991);
N. Read and G. Moore,
Prog. Theor. Phys. (Kyoto) Suppl. {\bf 107}, 157 (1992).

\bibitem{Radu08}
I.P. Radu, J.B. Miller, C.M. Marcus, M.A. Kastner, L.N. Pfeiffer,
and K.W. West, Science \textbf{320}, 899 (2008).

\bibitem{rmp}
Chetan Nayak, Steven H. Simon, Ady Stern, Michael Freedman, and Sankar Das Sarma,
Rev. Mod. Phys. {\bf 80}, 1083 (2008).

\bibitem{Read96}
N. Read and E. H. Rezayi,
Phys. Rev. B\textbf{54}, 16864 (1996).

\bibitem{Read99}
N. Read and E. H. Rezayi,
Phys. Rev. B\textbf{59}, 8084 (1999).

\bibitem{hr}
F. D. M. Haldane and E. H. Rezayi,
Phys. Rev. Lett. {\bf 60}, 956 (1988).

\bibitem{rg}
N. Read and D. Green,
Phys. Rev. B {\bf 61}, 10267 (2000).

\bibitem{nick}
N. Read,
Phys. Rev. B {\bf 79}, 045308 (2009);
''Quasiparticle spin from adiabatic transport in quantum Hall trial wavefunctions``,
e-print arXiv:0807.3107

\bibitem{steve}
S. H. Simon, E. H. Rezayi, N. R. Cooper, and I. Berdnikov,
Phys. Rev. B {\bf 75}, 075317 (2007).

\bibitem{HRsqueezed}
 E. H. Rezayi and F. D. M. Haldane,
Phys. Rev. B\textbf{50}, 17199 (1994).


\bibitem{FDM}
F. D. M. Haldane, chapter 8, p. 303, contribution in
\textit{The Quantum Hall Effet},
edited by R. E. Prange and S. M. Girvin (Springer Verlag, New York, 1990).

\bibitem{Jolad09}
S. Jolad and J. K. Jain,
Phys. Rev. Lett. {\bf 102}, 116801 (2009).

\bibitem{Jeon04}
G. S. Jeon, C.-C. Chang, and J. K. Jain,
Phys. Rev. B\textbf{69}, 241304(R)(2004);
Eur. Phys. J. B\textbf{55}, 271 (2007).

\bibitem{Barberan06}
N. Barberan, M. Lewenstein, K. Osterloh, and D. Dagnino,
Phys. Rev. A\textbf{73}, 063623 (2006).

\bibitem{Baksmaty07}
L. O. Baksmaty, C. Yannouleas, and U. Landman,
Phys. Rev. A\textbf{75}, 023620 (2007).


\bibitem{MacDoBook}
I. G. McDonald, {\it ''Symmetric Functions and Hall polynomials``}, 2nd edition,
Oxford University Press, New York (1995).

\bibitem{Feigin}
B. Feigin, M. Jimbo, T. Miwa, and E. Mukhin,
Int. Math. Res. Not. {\bf 2002}, 1223 (2002).

\bibitem{halperin}
B. I. Halperin,
Helv. Phys. Acta {\bf 56}, 75 (1983).

\bibitem{bh}
B. A. Bernevig and F. D. M. Haldane,
Phys. Rev. Lett. {\bf 100}, 246802 (2008);
{\textit ibid.} {\bf 101}, 246806 (2008);
{\textit ibid.} {\bf 102}, 066802 (2009);
Phys. Rev. B {\bf 77}, 184502 (2008).

\bibitem{rbh}
N. Regnault, B. A. Bernevig and F. D. M. Haldane,
''Topological Entanglement and Clustering of Jain Hierarchy States``,
e-print arXiv:0901.2121

\bibitem{mmnr}
M. Milovanovi\'{c} and N. Read,
Phys. Rev. B {\bf 53}, 13559 (1996).

\bibitem{gn1}
V. Gurarie, M. Flohr, and  C. Nayak,
Nucl. Phys. B {\bf 498}, 513 (1997).

\bibitem{sky}
We have used the notation of A. Seidel and K.
Yang, Phys. Rev. Lett. {\bf 101}, 036804 (2008), where this
configuration encodes one of the eight ground states of the (331)
case on the thin torus.

\bibitem{comment}
It is helpful to compare the configuration
(\ref{331root}) to the ones with non-Abelian quasiholes in the
polarized cases in Ref.~(\onlinecite{bh}).

\bibitem{bgs}
B. A. Bernevig, V. Gurarie, and S. H. Simon,
J. Phys. A: Math. Theor. \textbf{42}, 245206 (2009).

\bibitem{eddy}
E. Ardonne, Phys. Rev. Lett. {\bf 102}, 180401 (2009).

\bibitem{gin}
P. Ginsparg,
in {\em Fields, Strings and Critical Phenomena}
(Les Houches, Session XLIX, 1988) ed. by E. Br\'ezin and J. Zinn
Justin (1989).
(hep-th/9108028)

\bibitem{gn2}
The first application of this background charge representation in the
context of non-Abelian states was given in
V. Gurarie and C. Nayak,
Nucl. Phys. B {\bf 506}, 685 (1997).

\bibitem{df}
Vl. S. Dotsenko and V. A. Fateev,
Nucl. Phys. B {\bf 240}[FS12], 312 (1984);
Vl. S. Dotsenko and V. A. Fateev,
Nucl. Phys. B {\bf 251}[FS13], 691 (1985).

\bibitem{zub}
J. B. Zuber,
in {\em Fields, Strings and Critical Phenomena}
(Les Houches, Session XLIX, 1988) ed. by E. Br\'ezin and J. Zinn
Justin (1989);
P. di Francesco, H. Saleur, and J. B. Zuber,
J. Stat. Phys.{\bf 49}, 57 (1987).

\bibitem{car}
J. Cardy,
Lectures given at Les Houches Summer School on
{\em Exact Methods in Low-Dimensional Statistical Physics and
Quantum Computing}, July 2008,
arXiv:0807.3472

\bibitem{footnote}
For the
Coulomb gas representation of the two remaining primary fields in
${\cal M}(3,5)$ theory; $\psi$ and $\varphi$ with conformal weights
$h_{\psi} = \frac{3}{4}$ and $h_{\varphi} = \frac{1}{5}$, we have the
corresponding vertex operators with $\beta^{\psi}_{1,2} =
\frac{1}{\sqrt{15}}(1 \pm \frac{7}{2})$ and $\beta^{\varphi}_{1,2} =
\frac{1}{\sqrt{15}} (1 \pm 2)$ respectively.


\bibitem{bs}
A. Balatsky and M. Stone,
Phys. Rev. B {\bf 43}, 8038 (1991).

\bibitem{nrah}
N. Read,
Phys. Rev. Lett. {\bf 65}, 1502 (1990).

\bibitem{frie}
D. Friedan, E. Martinec, and S. Shenker,
Nucl. Phys. B{\bf 271}, 93 (1986).

\bibitem{mmnrss}
M. Milovanovi\'c and N. Read,
Phys. Rev. B{\bf 56}, 1461 (1997).

\bibitem{WenZee}
X. G. Wen and A. Zee,
Phys. Rev. Lett. {\bf 69}, 1811 (1992).

\bibitem{nonAjain} J.K. Jain, Phys. Rev. B{\bf 40}, 8079 (1989);
X. G. Wen, Phys. Rev. Lett. {\bf 66}, 802 (1991).

\bibitem{nregnault}
We thank N. Regnault for pointing out
this orbital singlet (L = 0) configuration to us.

\bibitem{jain}
J. K.  Jain,
Phys. Rev. Lett. {\bf 63}, 199 (1989);
Phys. Rev. B {\bf 40}, 8079 (1989);
ibid. {\bf 41}, 7653 (1990).




\end{thebibliography}
\end{document}